\documentclass[12pt]{article}
\usepackage{graphicx}

\newcommand{\1}{{I}}
\newcommand{\2}{{II}}
\newcommand{\dst}{\displaystyle}

\begin{document}

\title{Light curve and neutrino spectrum emitted during the collapse
of a nonrotating, supermassive star.}

\author{A.N. Baushev \thanks{Space Research Institute, Russian Academy of Sciences, Profsoyuznaya str. 84/32, Moscow, 117810 Russia\endgraf
Bogoliubov Laboratory of Theoretical Physics, Joint Institute for Nuclear Research;
141980 Dubna, Moscow Region, Russia\endgraf
Email: abaushev@mx.iki.rssi.ru}  \and G.S. Bisnovatyi-Kogan
\thanks{Space Research Institute, Russian Academy of Sciences, Profsoyuznaya str. 84/32, Moscow, 117810 Russia,
Email: gkogan@mx.iki.rssi.ru} }

\date{}

\maketitle

\begin{abstract}
The formation of a neutrino pulse emitted during the relativistic collapse of a spherical supermassive
star is considered. The free collapse of a body with uniform density in the absence of rotation
and with the free escape of the emitted neutrinos can be solved analytically by quadrature. The light
curve of the collapsing star and the spectrum of the emitted neutrinos at various times are calculated.
\end{abstract}

\section{INTRODUCTION}

The properties of a neutrino pulse arising during
the spherically symmetrical collapse of a supermassive
star into a black hole have been considered in a
number of studies, the earliest of which were carried
out nearly 40 years ago \cite{zeld63}.
It was shown in \cite{poduretz64} that, in
general, the asymptotic time dependence of the total
luminosity of the star has the form
$$ I \propto \exp \displaystyle{\left(-\frac{c t}{3
\sqrt{3} r_g}\right)} $$
The propagation of photons outside a nonrotating
collapsing star was considered in detail in \cite{amesthorn}.
The brightness distribution across the visible disk and
the time dependence of the stellar spectrum were
obtained for late stages of the collapse. The results
of numerical computations of the light curve and the
spectrum of a nonrotating, collapsing star are presented
by Shapiro \cite{shapiro89, shapiro96}, who computed the propagation
of photons inside the star in a diffusion approximation
taking into account general relativistic effects.
The propagation of photons to a distant observer is
also treated in detail in \cite{shapiro89}. Estimates of the parameters
of the spectrum and of the total intensity of a
neutrino pulse arising during the collapse of a supermassive
state were obtained in \cite{shi98}. The possibility of
detecting such a pulse using modern detectors was
analyzed, and the general background of neutrinos
arising as a result of explosions of supermassive stars
was estimated. The results of relativistic hydrodynamical
computations of the collapse of a nonrotating,
supermassive star are presented in \cite{linke01},
together with temperature and density profiles for the matter
inside the star at various times. However, the propagation
of the neutrinos was computed using a very
simple scheme; it was essentially assumed that the neutrinos
propagated instantaneously. In addition, all
general relativistic effects were neglected.

Thus, in spite of the abundance of publications
on the topic, the properties of neutrino pulses arising
during the collapse of supermassive stars are not fully
understood. As a rule, previous papers have either
considered very simple models for this phenomenon
that have admitted exact solutions (as in \cite{zeld63}), constructed
very detailed models that have required the
use of approximation computational methods, or have
neglected certain physically important effects. For
example, the propagation of the neutrinos is taken to
be instantaneous in \cite{linke01}, and gravitational forces are
not taken into account. Such an approach is clearly
not justified physically. During the collapse of a supermassive
star, the temperature of the matter grows
as the collapse proceeds; accordingly, the neutrinoloss
rate in a comoving frame also grows, reaching
its highest values in the late stages of the collapse.
The drop in the radiation intensity at the end of the
collapse is due exclusively to gravitational effects:
the gravitational redshift and time dilation. Thus, the
shape of the light curve is determined by the battle
between two competing processes: the increase in
the brightness of the star due to heating and the
decrease in its brightness due to gravitational effects.
In addition, the characteristic time scale for the collapse
$\displaystyle{\frac{r_g}{c}}$ is comparable to the time for a neutrino
to travel a distance equal to the radius of the star.
Under such conditions, the neglect of gravitational
effects and of the neutrino-propagation time is completely
unjustified. Ordinary (not supermassive) stars
were considered in \cite{shapiro96}; these computations included
all general relativistic effects, but the propagation of
the neutrinos inside the star was treated in an Eddington
approximation, which cannot be formulated unambiguously inside the star, where the effects of
sphericity are appreciable.

The aim of the current paper is to construct a
model for the neutrino emission radiated by a supermassive
star that can yield as accurate a solution
as possible, including all general relativistic effects,
while simultaneously providing a good description of
a real collapsing star. The light curve and spectra
obtained with this model should display all the main
properties characteristic of the light curves and spectra
of real collapsing supermassive stars, while the
simplicity of the model enables us to understand the
dependence of the parameters of the neutrino impulse
on the parameters of the problem. As we will show
below, by introducing a number of natural constraints
on the parameters of the problem, we can obtain a
solution by quadrature.

\section{DYNAMICS OF THE COLLAPSE}

We introduce a number of assumptions about the
physical parameters of the collapse that simplify the
computations. Most importantly, the star is taken to
be nonrotating and the distribution of matter to be
uniform at all times. In addition, we neglect completely
the influence of the pressure on the motion of
the matter; i.e., the dynamics of the collapse are taken
to be the same as for a sphere of dust. We also assume
that the matter is heated according to a power law as
a result of compression.
Our neglect of the pressure is a rather crude approximation
at the onset of the collapse, but the temperature
of the matter is still modest at this stage, and
few neutrinos are radiated. As the star approaches its
gravitational radius $r_g$, the influence of the pressure
tends to zero, so that it indeed becomes negligible.
As is shown in \cite{teorpol}, inside a homogeneous dust
sphere that is shrinking or expanding, it is possible to
introduce a comoving, synchronous coordinate system
in which ordinary three-dimensional space has
the same curvature everywhere, which depends only
on time. The character of this curvature (positive,
negative, zero) does not change in the course of the
collapse and is uniquely specified by the initial conditions
(initial density and velocity). We will take the
three-dimensional space inside the sphere to have
zero curvature, which appreciably simplifies the computations
without introducing large errors.

Let the initial radius of the star be $x_0$\footnote{Here and below, unless explicitly stated otherwise, we will
work with a physical system of units in which $c=1$ and
$r_g=\frac{2kM}{c^2}=1$, which specifies the units of velocity ($c$),
length ($r_g$), and time ($r_g/c$).}.
According to \cite{teorpol}, the metric inside the dust sphere can be written as
\begin{eqnarray}
ds^2=d\tau^2 - a^{2}(\tau)[dR^2 + R^2 (d\zeta^2+\sin^2\! \zeta\;
d\xi^2)] \label{a1}\\ \mbox{where} \quad
a(\tau)=\left(1-\frac{\tau}{\tau_{col}}\right)^{\frac{2}{3}}\label{a2}
\end{eqnarray}
Here $\tau_{col}$, is a parameter of the problem that is defined
below and the "spatial" coordinates ($R, \zeta, \xi$) are
analogous to the usual spherical coordinates in threedimensional
space.T he coefficient $a$ plays the role of
a scale factor: to find the physical distance between
two points in the metric (\ref{a1}), we must multiply the
"coordinate distance" [calculated in the coordinates  ($R, \zeta,
\xi$)) by $a$. In particular, the distance from a point
with radial coordinate $R$ to the center of the star is:
\begin{equation}
\label{1a2}
x=a R
\end{equation}
The coordinate system ($\tau, R, \zeta, \xi$) is comoving.
This means that each element of the collapsing matter
has some fixed $R, \zeta$ and $\xi$ that do not change in time.
In this sense, ($R, \zeta, \xi$) can be considered Lagrange
coordinates.

Note the full analogy between the solution for a
collapsing dust sphere considered here and the Friedmann
solution for a flat universe filled with homogeneous
dust, whose metric coincides fully with \ref{a1}. Therefore, the inner space of the dust sphere can be
considered part of a Friedmann universe; the motion
of the neutrinos within the dust sphere occurs precisely
as within a Friedmann universe.
In a comoving coordinate system, the radial coordinate
$R_s$ of the surface does not change and at any
time $\tau < \tau_{col}$ is equal to its initial value
\begin{equation}
\label{a3}
R_s=\frac{x_s}{a}=R_{s0}=\frac{x_{s0}}{a_0}
\end{equation}
In addition, it is straightforward to obtain the following
relation using the results of \cite{teorpol}:
\begin{equation}
\label{a4} 2 \pi k \varepsilon = \frac{1}{3\tau^{2}_{col} a^3}
\end{equation}
Here, $k$ is the gravitational constant and $\varepsilon$ is the density
of the stellar matter in its rest frame.Integrating $\varepsilon$
over the entire volume of the collapsing sphere yields
the total rest mass of the stellar matter. It is of interest
to determine how this quantity is related to the total
mass of the star measured by a distant stationary
observer. This latter mass consists of the rest mass
of the matter and the sum of its kinetic and potential
energy. As was noted above, we are considering the
collapse of a dust sphere within which the three-dimensional
curvature of space is zero. In this case,
the motion of dust matter toward the center is a free
fall with zero velocity at infinity. The sum of the kinetic
and potential energy of matter undergoing such a free
fall is zero. Therefore, the mass of the star is equal to
the rest mass of its matter, i.e.,
\begin{equation}
\label{a5}
M = \int \varepsilon  dV
\end{equation}
Hence, taking into account also the homogeneity of
the star, we have
\begin{equation}
\label{a6}
\varepsilon = \frac{3 M}{4 \pi x^3_s}
\end{equation}
Combining this relation with (\ref{a4}) yields
\begin{equation}
\label{a7} \frac{3 k M}{2 x^3_s} = \frac{1}{3 \tau^{2}_{col} a^3},
\quad \mbox{in other words, in accordance with (\ref{a3})} \quad k M = \frac{2
R^{3}_{s}}{9 \tau^{2}_{col}}.
\end{equation}
Since in our units $r_g = c = 1$, it follows from the
definition $r_g=\frac{2kM}{c^2}$ that
\begin{equation}
\label{a8}
k M = \frac{1}{2}
\end{equation}
which yields
\begin{eqnarray}
\frac{4 R^{3}_s}{9 \tau^{2}_{col}} = 1, \qquad \tau_{col} = \frac{2}{3}
R^{\frac{3}{2}}_s \label{a9}
\end{eqnarray}
We will adopt the time $\tau = 0$ as the initial time.
Then, $a_0 = 1$ and
\begin{equation}
\label{a10} x_{s0} = R_{s0}=R_s.
\end{equation}
Having specified the initial radius of the star $x_{s0}$, we can use (\ref{a9})
to determine the collapse parameter $\tau_{col}$. The collapse of a uniform sphere with a flat internal
space involves two independent parameters: the
Lagrange radius of the surface $R_s$  and the mass of
the star. The quantity $\tau_{col} -\tau_0$ is uniquely determined
by these two parameters, and variations in $\tau_0$ only
shift the zero time. In the case of arbitrary three-dimensional
curvature, we can arbitrarily specify the
mass, initial radius, and initial collapse velocity of the
star. If we consider a collapse with zero curvature,
this is not possible, and the initial collapse velocity
always differs from zero for a finite initial radius. This
is automatically determined from the condition of zero
curvature and turns out to be equal to the velocity for
free fall from infinity for the radius $x_{s0}$. In the interest
of conciseness, we will denote the initial radius of the
star simply as $x_0$.

\section{COMPUTATION OF THE RADIATION SPECTRUM IN A COMOVING FRAME}
We will assume that each matter element emits
neutrinos isotropically in a comoving frame and that
the intensity and spectrum of the emitted neutrinos
depend on the temperature and density of the matter.
In our analysis, the dynamics of the collapse corresponds
to the motion of the dustlike material. If the
law for variations of the density and the parameters
$M$ and $R_s$ is known, the variations of the temperature
in the compressed matter can be found from the
solution of the energy equation for a specified initial
entropy using the known rate of neutrino emission
as a function of the temperature and density. Below,
we use a simplified approach in order to obtain
an analytical solution, in which we approximate the
time dependence of the temperature with a power-law
dependence of the temperature on the density. The
variations of the density of the matter $\varepsilon(\tau)$ are determined
in accordance with (\ref{a3}) and (\ref{a6}) by the single
function $a(\tau)$. To find the observed light curve and
spectrum of the neutrinos, we must first determine the
propagation toward the observer of a neutrino signal
arising within a uniform collapsing body of a specified
mass $M$.

We will suppose that
\begin{equation}
\label{a11} f(a,q) dV d^3p d\tau
\end{equation}
neutrinos are emitted in an element of physical\footnote{Not coordinate.} volume
$dV$ in an element of physical phase space $d^3p$
in an interval of proper time $d\tau$ in the element's rest
frame, where $f(a,q)$ is some function whose form will
be determined below.

We introduce the neutrino distribution\footnote{We will call a quantity defined
in this way a physical distribution function.} function
$N$ such that the quantity $N d^3p dV$ specifies the
number of neutrinos in an element of physical volume
$dV$ in an element of physical phase space $d^3p$. In
the spherically symmetric case with isotropic neutrino
emission in the comoving frame, $N$ depends on only
four quantities: the radial coordinate $R$, the angle
between the neutrino trajectory and the direction from
the center of the star $\vartheta$, the neutrino energy $q$, and
time $\tau$. In this case\footnote{We also apply here the condition of zero three-dimensional
curvature of the space inside the star.}, the kinetic equation can be
written in the form \cite{lindquist66}
\begin{equation}
\label{a12} \cos\vartheta\frac{\partial N}{\partial R} -
\frac{\sin\vartheta}{R}\frac{\partial N}{\partial \vartheta} +
\frac{2 q}{3 \tau_{col} \sqrt{a}}\frac{\partial N}{\partial q} -
\frac{2 \sqrt{a}}{3 \tau_{col}}\frac{\partial N}{\partial a}= a f(a,q)
\end{equation}
Here, the parameter $a$, which depends on time in
accordance with (\ref{a2}), is used in place of the time $\tau$. The
characteristic system for this equation has the form
\begin{eqnarray}
\frac{dR}{\cos\vartheta} = dy \label{a13}\\
-\frac{R d\vartheta}{\sin\vartheta} = dy \label{a14}\\
\frac{3 \tau_{col} \sqrt{a} dq}{2 q} = dy \label{a15}\\
-\frac{3 \tau_{col} da}{2 \sqrt{a}} = dy \label{a16}
\end{eqnarray}
The solutions of this system determine the trajectories
of individual neutrinos.Combining pairwise the
equations for characteristic system (\ref{a15}) and (\ref{a16}), and
also (\ref{a13}) and (\ref{a14}), we obtain the two first integrals:
\begin{eqnarray}
A = a q \label{a17}\\
B = R \sin\vartheta \label{a18}
\end{eqnarray}
Combining (\ref{a14}) and (\ref{a16}) and using
(\ref{a18}), we obtain
the third, last, integral of the system (\ref{a13})-(\ref{a16}).
\begin{equation}
\label{a21} C=3\tau_{col}\sqrt{a}+R\cos \vartheta
\end{equation}
The physical meaning of these quantities can be
elucidated based on the analogy between the solution
for a collapsing dust sphere and the solution for a
Friedmann universe. As was noted above, the motion
of a neutrino is absolutely identical in these two cases;
in particular, all three of the first integrals of system
(\ref{a13})-(\ref{a16}) are also valid in a Friedmann universe. The
presence of integral (\ref{a17}), i.e., the fact that the product
$a\cdot q$ is constant, leads to the appearance of a "redshift"
in a Friedmann universe. As the parameter $a$
increases (corresponding to the expansion of space),
the energy of a neutrino along its trajectory decreases.
In contrast, in our case we have a compression of a
sphere, so that $a$ decreases and the neutrino energy
increases. Thus, we can think of our case as giving
rise to a "violet shift".

Supermassive stars with masses exceeding
$\sim 10^5 M_\odot$ have small optical depths to the absorption
of neutrinos, which can be taken to propagate freely.
Taking the radius of the star to be $R = n R_g =
2 n G M / c^2$, we obtain expressions for the mean density,
$$
\rho=\frac{3 M}{4 \pi R^3}=\frac{1.8\cdot 10^{16}}{n^3}\left(\frac{M_\odot}{M}\right) g/cm^3
$$
and the optical depth to neutrino absorption,
$$
\tau_\nu=10^{-44}\rho R=\frac{33}{n^2}\frac{M_\odot}{M}\ll 1
$$
for $n\simeq 3$, $M=10^5 M_\odot$.

Let us elucidate the qualitative form of the characteristics
determining system (\ref{a13})-(\ref{a16}). The motion
of photons and neutrinos is linear in a Friedmann
universe. Indeed, integral (\ref{a18}) shows that a neutrino
moves along straight lines in the ($\tau, R, \zeta, \xi$) coordinate
frame. In addition, since this frame is comoving,
the collapsing star can be represented at any time
in this frame as a sphere with a constant radius $R_s$.
The trajectory described by (\ref{a13})-(\ref{a16}) proves to be a
chord of this sphere. In accordance with the definition
of the angle $\theta$, the trajectory enters the sphere at
some angle $\vartheta > \frac{\pi}{2}$ and leaves at an angle $\vartheta < \frac{\pi}{2}$.
Proceeding from this property of the characteristic, it
is straightforward to impose the appropriate boundary
conditions for (\ref{a12}). For this, it is sufficient to specify
the value of the function $N$ at the surface of the star
for angles $\vartheta > \frac{\pi}{2}$. We will assume that no neutrinos
impact the star from outside, in which case
\begin{equation}
\label{n6} \left. N\right|_{R=R_s}=0 \quad \mbox{ïðè}\quad
\vartheta > \frac{\pi}{2}
\end{equation}
This is the boundary condition for (\ref{a12}).

We will denote the initial point of the characteristic
and all related quantities with the subscript \1 and the
final point and related quantities with the subscript \2.
Since both points are located on the surface of the
collapsing sphere,
\begin{equation}
\label{a22}
R_\1=R_\2=R_s
\end{equation}
Hence, we obtain using (\ref{a18}):
\begin{equation}
\label{a23}
\sin \vartheta_\1=\sin \vartheta_\2
\end{equation}
Since $\vartheta_\1 > \frac{\pi}{2}, \vartheta_\2 < \frac{\pi}{2}$,
\begin{equation}
\label{a24}
\cos \vartheta_\1=-\cos \vartheta_\2
\end{equation}
In addition, it follows from the existence of the integral (\ref{a17}) that
\begin{equation}
\label{a25}
a_\1 q_\1=a_\2 q_\2 \quad\mbox{ò.å.} \quad a_\1=a_\2\frac{q_\2}{q_\1}
\end{equation}
Finally, writing integral (\ref{a21}), we obtain
\begin{equation}
\label{a26} 3\tau_{col}\sqrt{a_\1}+R_\1\cos \vartheta_\1=
3\tau_{col}\sqrt{a_\2}+R_\2\cos \vartheta_\2
\end{equation}
Using relations (\ref{a24}) and (\ref{a25}), we obtain
\begin{equation}
\label{a27} \sqrt{a_\2\frac{q_\2}{q_\1}}=\sqrt{a_\2}+\frac{2
R_s}{3\tau_{col}}\cos \vartheta_\2
\end{equation}
\begin{equation}
\label{a28}  q_\1=\frac{q_\2}{\left(\dst{1+\dst{\frac{2
R_s}{\dst{3\tau_{col}\sqrt{a_\2}}}}\cos\vartheta_\2}\right)^2}
\end{equation}
Let us turn to the solution of (\ref{a12}) itself. We first take
the neutrino-emission function (\ref{a11}) to be
\begin{equation}
\label{a19}
f(a,q) = \lambda a^{\beta} \delta(q-q_0)
\end{equation}
i.e., each element of matter in the proper frame radiates
monochromatic neutrinos with energy $q_0$. Here, $\lambda$
and $\beta$ are numerical parameters. Since (\ref{a12}) is linear,
the resulting solution can easily be generalized
to the case of a more complex function $f(a,q)$. We
will discuss the form of this function for a real system
and possible values of the power-law index $\beta$ in more
detail below.

For a neutrino-emission function of the form (\ref{a19}), Eq. (\ref{a12}) can be written in the form
\begin{equation}
\label{a20} \cos\vartheta\frac{\partial N}{\partial R} -
\frac{\sin\vartheta}{R}\frac{\partial N}{\partial \vartheta} +
\frac{2 q}{3 \tau_{col} \sqrt{a}}\frac{\partial N}{\partial q} -
\frac{2 \sqrt{a}}{3 \tau_{col}}\frac{\partial N}{\partial a} = \lambda
a^{\beta+1} \delta(q-q_0)
\end{equation}
To solve this equation, we must integrate the righthand
side along the characteristic.Using boundary
condition (\ref{n6}), we obtain for the surface of the collapsing sphere:
\begin{equation}
\label{a29}
N=\int\limits_\1^\2\lambda a^{\beta+1} \delta(q-q_0) dy
\end{equation}
We obtain using (\ref{a17})-(\ref{a28})
\begin{equation}
\label{a30} N=\int\limits_\1^\2 \frac{3}{2}\tau_{col}\lambda
\frac{A^{\beta+\frac{3}{2}}}{q^{\beta+\frac{5}{2}}} \delta(q-q_0)
dq= \frac{3}{2}\tau_{col}\lambda
\frac{A^{\beta+\frac{3}{2}}}{q_0^{\beta+\frac{5}{2}}}
\left(\Xi[q_\2-q_0]-\Xi[q_\1-q_0]\right)
\end{equation}
Here, the function $\Xi$ is defined to be
$$
\Xi[x]=
\left\{
\begin{array}{rcl}
0 \qquad {\textrm {ïðè}}\qquad x<0,\\
1 \qquad {\textrm {ïðè}}\qquad x>0.\\
\end{array}
\right.
$$
Note that quantities relating to the end of the characteristic
are those that are measured by an observer located
on the surface of the collapsing star. Therefore,
we can omit the subscript "\2". Substituting relations
(\ref{a17}) and (\ref{a28}) into (\ref{a30}), we finally obtain
\begin{equation}
\label{w30}
N=\frac{3}{2}\tau_{col}\lambda
\frac{a^{\beta+\frac{3}{2}}
q^{\beta+\frac{3}{2}}}{q_0^{\beta+\frac{5}{2}}}
\left(\Xi[q-q_0]
-\Xi\biggl[
q-\left(1+
\dst{
\frac{2R_s}{\dst{3\tau_{col}\sqrt{a}}}}\right)^2 q_0
\biggr]\right)
\end{equation}
This formula specifies the neutrino distribution function
measured by an observer located on the surface
of the star and moving with this surface.

\section{PROPAGATION OF NEUTRINOS
FROM THE STAR TO AN OBSERVER}
Let us now consider the motion of the neutrinos
outside the star.The gravitational field here is a
Schwarzschild field \cite{relastr}; i.e., it has the metric
\begin{equation}
ds^2=\left(1-\frac{1}{r}\right)dt^2 - \left(1-\frac{1}{r}\right)^{-1} dr^2
- r^2 (d\theta^2+\sin^2\! \theta\; d\phi^2) \label{b1}
\end{equation}
The neutrino distribution function $N$ in such a field
in the spherically symmetrical case depends on four
quantities: the angle between the neutrino trajectory
and the direction from the center of the star $\theta$, the
neutrino energy $w$, the radius $r$, and the time $t$.
We can find N by solving the kinetic equation, for
which the boundary conditions must be obtained using
(\ref{w30}). Let us determine the relationship between
physical quantities measured on the surface of the
star in the Friedmann and Schwarzschild reference
frames.Loc al observers in the Schwarzschild frame
are stationary, while local observers in the Friedmann
frame move along with the stellar material. Therefore,
physical quantities measured by such observers at
some time at the same point on the stellar surface
will be related by the usual Lorentz transformation.
The velocity in this transformation is the physical
velocity of the stellar surface measured by a local
Schwarzschild observer, which will be equal to \cite{relastr}
\begin{equation}
\label{b2}
V=\frac{1}{\sqrt{h}}
\end{equation}
where $h$ is the current Schwarzschild radius of the
star.

The circumference of the stellar surface in the Friedmann frame is
$2 \pi x$, and in the Schwarzschild
frame, $2 \pi h$. Since dimensions perpendicular to the
motion are not changed by the Lorentz transformation,
$2 \pi x = 2 \pi h$, i. e., $x=h$. Thus, the Schwarzschild
radius of the star is equal to $x$. In particular, (\ref{b2}) can
be rewritten as
\begin{equation}
\label{b3}
V=\frac{1}{\sqrt{x}}
\end{equation}
As is shown in \cite{teorpol}, the physical distribution function
of the particles is Lorentz
invariant. Therefore, to recalculate function (\ref{w30}) in
the Schwarzschild frame, we can simply express
the variables $(q, \cos\vartheta)$ in terms of the quantities
$(w_s, \cos\theta_s)$ in the Schwarzschild frame\footnote{Here and below, the subscript $s$ denotes variables in the
Schwarzschild frame specified on the surface of the collapsing
sphere.} using the Lorentz transformation and the velocity determined by (\ref{b3}).
Using the results of \cite{teorpol}, it is straightforward to obtain the relations:
\begin{eqnarray}
q=w_s \frac{1+V\cos\theta_s}{\sqrt{1-V^2}}=
w_s \frac{\sqrt{x}+\cos\theta_s}{\sqrt{x-1}} \label{b4}\\
\cos\vartheta=\frac{\cos\theta_s +V}{1+V\cos\theta_s}=
\frac{\sqrt{x}\cos\theta_s +1}{\sqrt{x}+\cos\theta_s} \label{b5}
\end{eqnarray}
Substituting (\ref{b4}) and (\ref{b5}) into (\ref{w30}),
we finally obtain
\begin{eqnarray}
\left. N\right|_{r=x}=\frac{\lambda}{x^{\beta}_0}
\frac{x^{\beta+\frac{3}{2}}}{q^{\beta+\frac{5}{2}}_0} \left(w_s
\frac{\sqrt{x}+\cos\theta_s}{\sqrt{x-1}}
\right)^{\beta+\frac{3}{2}} \times \nonumber \\ \left(\Xi\left[w_s
 -\left(\frac{\sqrt{x-1}}{\sqrt{x}+\cos\theta_s}\right)q_0 \right] -
\Xi\left[w_s-\left(\frac{\sqrt{x-1}}{\sqrt{x}+\cos\theta_s}\right)\left(1+
\frac{\sqrt{x}\cos\theta_s +1}{\sqrt{x}\cos\theta_s+x}
\right)^2 q_0\right] \right) \label{b6}
\end{eqnarray}
Thus, we have derived the physical neutrino distribution
function on the stellar surface in the Schwarzschild
frame.T o fully specify the boundary conditions,
we must determine an expression describing the motion
of this surface.T he equation for radial free fall
of a body in a Schwarzschild field is such that the
velocity of the body is zero at infinity \cite{relastr} (as was
indicated above, the surface of the dust sphere moves
in precisely this way) and has the form
\begin{equation}
\label{b7}
t=\ln \left(\frac{\sqrt{x}+1}{\sqrt{x}-1}\right)-\frac{2}{3} x^{\frac{3}{2}}
-\sqrt{x}+D
\end{equation}
Here, $D$ is the constant of integration.By specifying
this constant, we determine the zero time in the
Schwarzschild frame. Setting $D$ equal to zero, we
obtain
\begin{equation}
\label{b8}
t=\ln \left(\frac{\sqrt{x}+1}{\sqrt{x}-1}\right)-\frac{2}{3} x^{\frac{3}{2}}
-\sqrt{x}
\end{equation}
This formula describes the motion of the stellar surface.
Thus, we have fully specified the boundary conditions
for the kinetic equation in the Schwarzschild
metric.The kinetic equation for such a field was obtained
in \cite{poduretz64}, where it was shown that, in this case, the
integrals of motion have the form
\begin{eqnarray}
\rho=\frac{r \sqrt{r}}{\sqrt{r-1}}\sin\theta  \label{b9}\\
\Omega=w \left(1-\frac{1}{r}\right)^\frac{1}{2} \label{b10}\\
\xi=t-\int\limits_{x(\xi)}^r\frac{z^2 \sqrt{z}}{(z-1)
\sqrt{z^3-\rho^2 (z-1)}}\, dz \label{b11}
\end{eqnarray}
The first of these quantities specifies the impact parameter
in the Schwarzschild field, and the second determines the redshift for the motion
of a neutrino in such a field. The third integral is essentially the time
when the neutrino is emitted from the stellar surface.
However, it will be more convenient for us to use in
place of the time when the neutrino is emitted, $\xi$, the
radius of the star at this time, $x$. It is obvious that
these quantities are unambiguously related to each
other in accordance with (\ref{b8}):
\begin{equation}
\label{b12}
\xi=\ln \left(\frac{\sqrt{x}+1}{\sqrt{x}-1}\right)-\frac{2}{3} x^{\frac{3}{2}}
-\sqrt{x}
\end{equation}
Let us now express the quantities $\cos\theta_s$ and $w_s$
in (\ref{b6}) in terms of the integrals of motion $\Omega$, $\rho$ and $x$.
We substitute the values of variables corresponding
to the stellar surface into formulas (\ref{b9})-(\ref{b11}) (in
particular, we must set $r=x$). This yields

\begin{eqnarray}
\rho=\frac{x \sqrt{x}}{\sqrt{x-1}}\sin\theta_s \label{b13}\\
\Omega=w_s \left(1-\frac{1}{x}\right)^\frac{1}{2} \label{b14}
\end{eqnarray}
Hence\footnote{Since the surface of the collapsing star is moving, some neutrinos
emitted from the surface have $\dst{\theta > \frac{\pi}{2}}$ at the time they
are emitted (as a consequence of the aberration of light). We
do not take these into account when deriving (\ref{b16}). However,
the fraction of such particles among those reaching a distant
observer will be very small, so that we are quite justified in
neglecting them.},
\begin{eqnarray}
w_s=\sqrt{\dst{\frac{x}{x-1}}}\Omega \label{b15}\\
\cos\theta_s=\frac{\sqrt{x^3-\rho^2 (x-1)}}{x\sqrt{x}} \label{b16}
\end{eqnarray}
Substituting the resulting expressions into formula
(\ref{b6}) for the neutrino distribution function at the stellar
surface,
\begin{eqnarray}
N=\frac{2 \lambda}{x^{\beta}_0} \frac{x^{\beta+\frac{3}{2}}
\Omega^{\beta+\frac{3}{2}}}{q^{\beta+\frac{5}{2}}_0
m^{\beta+\frac{3}{2}}}
\left(\Xi\left[\Omega - m q_0 \right] - \Xi\left[\Omega - g^2 m
q_0\right] \right) \label{n2}
\end{eqnarray}
Here, we have introduced the notation
\begin{eqnarray}
 u=\sqrt{x^3-\rho^2 (x-1)}; \quad m=\frac{x (x-1)}{x^2
+u}; \quad
\displaystyle{g=1+\frac{x+u}{x^2+u}}\label{n1}
\end{eqnarray}
Since the space outside the star does not contain
any neutrino sources, the the distribution function is
constant along the characteristics. Therefore, the resulting
expression specifies the neutrino distribution
at an arbitrary point outside the star.

However, we are ultimately interested in the parameters
of a neutrino pulse that is measured by an
observer infinitely distant from the collapsing star. Let
the distance from the observer to the star be $d$ $(d\gg 1)$.
Thus, we will be interested in the behavior of the
integrals (\ref{b9})-(\ref{b11}) for $r=d$, i. e., $r\gg 1$. The first two
integrals, ((\ref{b9}) and (\ref{b10})), acquire the form
\begin{eqnarray}
\rho=d \sin\theta \label{b27}\\
\Omega=w \label{b28}
\end{eqnarray}
As is noted above, we use $x$ in place of the integral
$\xi$ from system (\ref{b9})-(\ref{b11}); these quantities are unambiguously
related by (\ref{b12}). We substitute this relation
into (\ref{b11}) and express the time $t$ when the neutrino
reaches the radius $r$ in terms of $x$ and $\rho$.
\begin{equation}
\label{b29}
t=\ln \left(\frac{\sqrt{x}+1}{\sqrt{x}-1}\right)-\frac{2}{3} x^{\frac{3}{2}}
-2\sqrt{x}+\int\limits_x^r\frac{z^2 \sqrt{z}}{(z-1)
\sqrt{z^3-\rho^2 (z-1)}}\, dz
\end{equation}
We transform the integral in this formula by substituting $r$ with $d$.
\begin{eqnarray}
\int\limits_x^d\frac{z^2 \sqrt{z}}{(z-1) \sqrt{z^3-\rho^2 (z-1)}}\, dz =
d+\ln (d-1)-x-\ln (x-1) + \nonumber\\
+ \int\limits_x^d\frac{z}{z-1}\left[\frac{z \sqrt{z}}{\sqrt{z^3-\rho^2
(z-1)}}-1\right]\, dz \label{b30}
\end{eqnarray}
The last integral in this expression converges as
$d\to \infty$. Therefore, it is natural to set the upper limit
equal to infinity. However, we can see that the initial
integral of (\ref{b30}) diverges. To elucidate the physical
meaning of this fact, we substitute relation (\ref{b30}) into
the initial formula (\ref{b29}):
\begin{eqnarray}
t=d+\ln (d-1)-2\ln (\sqrt{x}-1)-\frac{2}{3}
x^{\frac{3}{2}}-2\sqrt{x}-x + \nonumber\\ +
\int\limits_x^{\infty}\frac{z}{z-1}\left[\frac{z
\sqrt{z}}{\sqrt{z^3-\rho^2 (z-1)}}-1\right]\, dz \label{b31}
\end{eqnarray}
We can see that the divergence of (\ref{b30}) leads to an unlimited
growth in the arrival time of the neutrino to the
observer as the distance to the observer approaches
infinity.A similar behavior shown by function (\ref{b29}) is
quite natural. However, it is clear that such a dependence
for the neutrino arrival time on the distance
to the observer has absolutely no effect on the light
curve or the observed spectrum; it only affects the
arrival time of the neutrino pulse. Therefore, we can
shift the zero time for an infinitely distant observer by
$d+\ln (d-1)$; i.e., we make the substitution
\begin{eqnarray}
\label{b133}
t^{*}=t-d-\ln (d-1)
\end{eqnarray}
Substituting this value into (\ref{b29}), we obtain:
\begin{eqnarray}
t^{*}=-2\ln (\sqrt{x}-1)-\frac{2}{3} x^{\frac{3}{2}}-2\sqrt{x}-x
+ \nonumber\\
+ \int\limits_x^{\infty}\frac{z}{z-1}\left[\frac{z \sqrt{z}}{\sqrt{z^3-\rho^2
(z-1)}}-1\right]\, dz \label{b32}
\end{eqnarray}
This expression relates the parameters $t^{*}$, $x$ and $\rho$ of
some phase trajectory. It can be considered a specification
of the integral of motion $x$ in the form of
an implicit function $x(\rho,t^{*})$ of the impact parameter
$\rho$ and the time $t^{*}$ of the arrival of the neutrino to
a distant observer. Everywhere below, by $x$, we will
mean precisely this function. Note that, since $t^{*}$ is a
function only of $x$ and $\rho$, i. e., of integrals of the motion,
it is also an integral of the motion.

Now, using (\ref{b27}) and (\ref{b28}), we can readily find
the number of neutrinos with energies in the interval
$[w, w+dw]$ passing through a sphere of radius $d$ over
a time $dt^{*}$. This quantity can be expressed in terms of
the distribution function $N$ as follows:
\begin{equation}
\label{n3} \Phi_0(w,t^{*})=8 \pi^2 w^2 \int\,\rho
N(w,\rho,x(\rho,t^{*}))\; d\,\rho
\end{equation}
where the integration is over all $\rho$ values admissible
for a given $t^{*}$ and $w$. Here, we have taken into account
the equality of $\Omega$ and $w$ that follows from (\ref{b28}).
Having integrated (\ref{n3}) over the energy, we can
obtain an expression for the light curve:
\begin{equation}
\label{n4} E_0(t^{*})=8 \pi^2 \int\!\!\int \rho w^3
N(w,\rho,x(\rho,t^{*}))\; d\rho dw
\end{equation}
We now substitute in the resulting expressions the
distribution function (\ref{n2}) and transform them into the form
\begin{equation}
\label{n5} \Phi_0(w,t^{*})= 8 \pi^2 \lambda \frac{q_0}{x^{\beta}_0}
\left(\frac{w}{q_0}\right)^{\beta+\frac{7}{2}}\int\rho
\left(\frac{x}{m}\right)^{\beta+\frac{3}{2}} \left[\Xi(w-m
q_0)-\Xi(w-g^2 m q_0)\right]\; d\rho
\end{equation}
\begin{eqnarray}
\label{n10} E_0(t^{*})=8 \pi^2 \lambda \frac{q^{2}_0}{x^{\beta}_0}
\int\!\!\int\rho
\left(\frac{w}{q_0}\right)^{\beta+\frac{9}{2}}
\left(\frac{x}{m}\right)^{\beta+\frac{3}{2}} \left[\Xi(w-m
q_0)-\Xi(w-g^2 m q_0)\right]\; d\rho dw = \nonumber\\
=\frac{8 \pi^2 \lambda}{\left(\beta+\dst{\frac{11}{2}}\right)} \frac{q^{3}_0}{x^{\beta}_0}
\int\rho m^4 x^{\beta+\dst{\frac{3}{2}}}  (g^{2\beta+11}-1) \; d\rho
\end{eqnarray}
In all these formulas, integration over $\rho$ is understood
to mean integration over all values of $\rho$ for which there
exists a solution to (\ref{b32}) for a given $t^{*}$. The resulting
formulas specify the desired analytical solution for the
source function of the form (\ref{a19}).

\section{APPLICATION OF THE MODEL
TO THE COLLAPSE OF A REAL STAR}
We can use our model to compute the light curve
and spectrum of a real collapsing star.W e take the
mass of the star to be $\dst{M=10^5 M_\odot}$. Since the intensity
of the neutrino emission by the stellar matter
grows with temperature, which, in turn, grows as the
star is compressed, it is expedient to consider only
the last stages of the collapse, when the vast majority
of neutrinos are emitted. We therefore take the initial
radius to be $x_0 = 10$.

Let us now consider the heating of the stellar matter
during the compression. In our model, this matter
is taken to be dust, which, of course, does not change
its temperature as it is compressed. Therefore, the law
describing the temperature variations of the matter
must be specified separately. We will assume that the
matter is heated like an ideal gas during an adiabatic
compression:
\begin{equation}
\label{n7} \frac{T}{\varepsilon^{\gamma-1}}=\mbox{\it const}
\end{equation}
Here, $T$ is the temperature of the matter and $\varepsilon$ is its
density. For a relativistic gas (radiation), $\displaystyle{\gamma=\frac{4}{3}}$. In
addition, it follows from (\ref{a4}) that $\varepsilon\propto a^{-3}$. Therefore,
we can rewrite formula (\ref{n7}) in the form
\begin{equation}
\label{n8} a T=\mbox{\it const}
\end{equation}
Denoting the initial temperature of the matter $T_0$
(and using the fact that $a_0 = 1$), we finally obtain the
formula
\begin{equation}
\label{n9} T=\frac{T_0}{a}
\end{equation}
The parameters for which a star loses its stability were
derived in \cite{bk68}, in a study of the evolution of massive
stars. If we suppose that the subsequent compression
is adiabatic and that the influence of the pressure can
be neglected, we can use the parameters presented
in \cite{bk68} to estimate the density and temperature when
the radius of the collapsing star is equal to $x_0$. We
obtain $\dst{\varepsilon_0=2\cdot 10^3
\frac{\mbox{g}}{\mbox{cm}}}$, $T_0=3\cdot 10^9 K$.

Let us now consider the physical processes occurring
up to the radiation of neutrinos by the heated
matter. As was shown in \cite{schinder87,itoh89}, the main mechanism
for the radiation of neutrinos during the collapse
of a supermassive star is the annihilation of electron–
positron pairs. In this case, the energy radiated per
unit volume of matter per unit time is specified by the
very simple formula \cite{shi98}\footnote{Expressions (\ref{n12}), (\ref{q2}) and (\ref{n15})
are written in the cgs system.}
\begin{equation}
\label{n12} Q\simeq 4\cdot 10^{-66} T^9\quad
\left(\frac{\mbox{erg}}{\mbox{cm}^3 \; \mbox{s}} \right)
\end{equation}
where the temperature is in Kelvin. Hence, $\beta = -9$.
When finding the light curve, the energy distribution
of the neutrinos is not important and we can take all
the emitted neutrinos to have the same energy, $q_0$. In
this case, the function $f(q,a)$ in (\ref{a11}) corresponding
to the radiation intensity (\ref{n12}) and the parameters of
the star presented above is given by
\begin{equation}
\label{q1}
\tilde f(a,q)=\frac{2\cdot 10^{29}}{q_0^3} a^{\beta} \delta(q-q_0)
\end{equation}
We thus obtain using (\ref{n10})\footnote{Note that, during the calculations, all quantities
in (\ref{q2}) and (\ref{n15}) must be taken in the system of units in which $c=1$ and
$c=1$ è $r_g=\frac{2kM}{c^2}=1$. The transition to cgs units is carried out
via the numerical coefficient preceding the formula.}:
\begin{equation}
\label{q2} E(t^{*})= 4 \cdot 10^{60}  \int\frac{\rho
m^4}{x^{\dst \frac{15}{2}}} (1-g^{-7}) d\rho \quad
\left(\frac{\mbox{erg}}{\mbox{s}} \right)
\end{equation}
This formula specifies the desired light curve.
\begin{figure}
\vspace*{-1cm}
\resizebox{0.75\hsize}{!}{\includegraphics[angle=270]{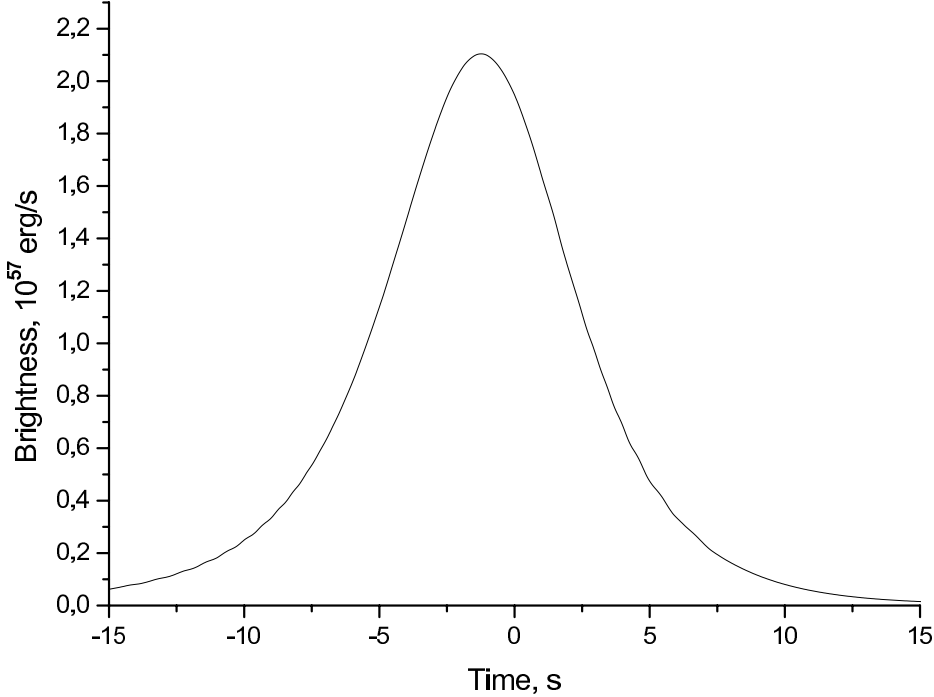}}
\caption{Neutrino light curve.The time in the plots is the
time $t^{*}$ for an infinitely distant observer whose zero time
has been shifted in accordance with (\ref{b133}). For more detail
about the choice of zero time for a distant observer, see
the text following formula (\ref{b28}).}
\label{fig1}
\end{figure}

Let us now consider the spectrum of the neutrinos.
The spectrum of neutrinos generated during
pair annihilations was calculated using Monte Carlo
simulations in \cite{shi98}; under the conditions of interest to
us, the shape of this spectrum is approximated well by
the formula
\begin{equation}
\label{n111} L(q)\propto\frac{\left(\dst{\frac{q}{T}}\right)^2}{1+
\exp{\left(\dst{\frac{q}{1,6 T}}-2\right)}}
\end{equation}
where $L$ is the energy radiated per unit volume of
matter per unit time in a unit interval of energy. We
can see that the shape of the spectrum depends on
the temperature. However, we will assume that temperature
variations affect only the total intensity of
the radiation, leaving the spectrum unchanged; i.e.,
the spectrum's shape will correspond to that for some
fixed temperature $E_0$:
\begin{equation}
\label{n11}  L(q)\propto\frac{\left(\dst{\frac{q}{E_0}}\right)^2}{1+
\exp{\left(\dst{\frac{q}{1,6 E_0}}-2\right)}}
\end{equation}
It is reasonable to choose the parameter $E_0$ so that it
is approximately equal to the temperature of the matter
when the luminosity of the star is maximum. We
will take this parameter to be equal to the temperature
of the matter when $x\simeq 1.43 r_g$; i. e., $E_0=7 T_0$ in our case.

\begin{figure}
\vspace*{-1cm}
\resizebox{0.75\hsize}{!}{\includegraphics[angle=270]{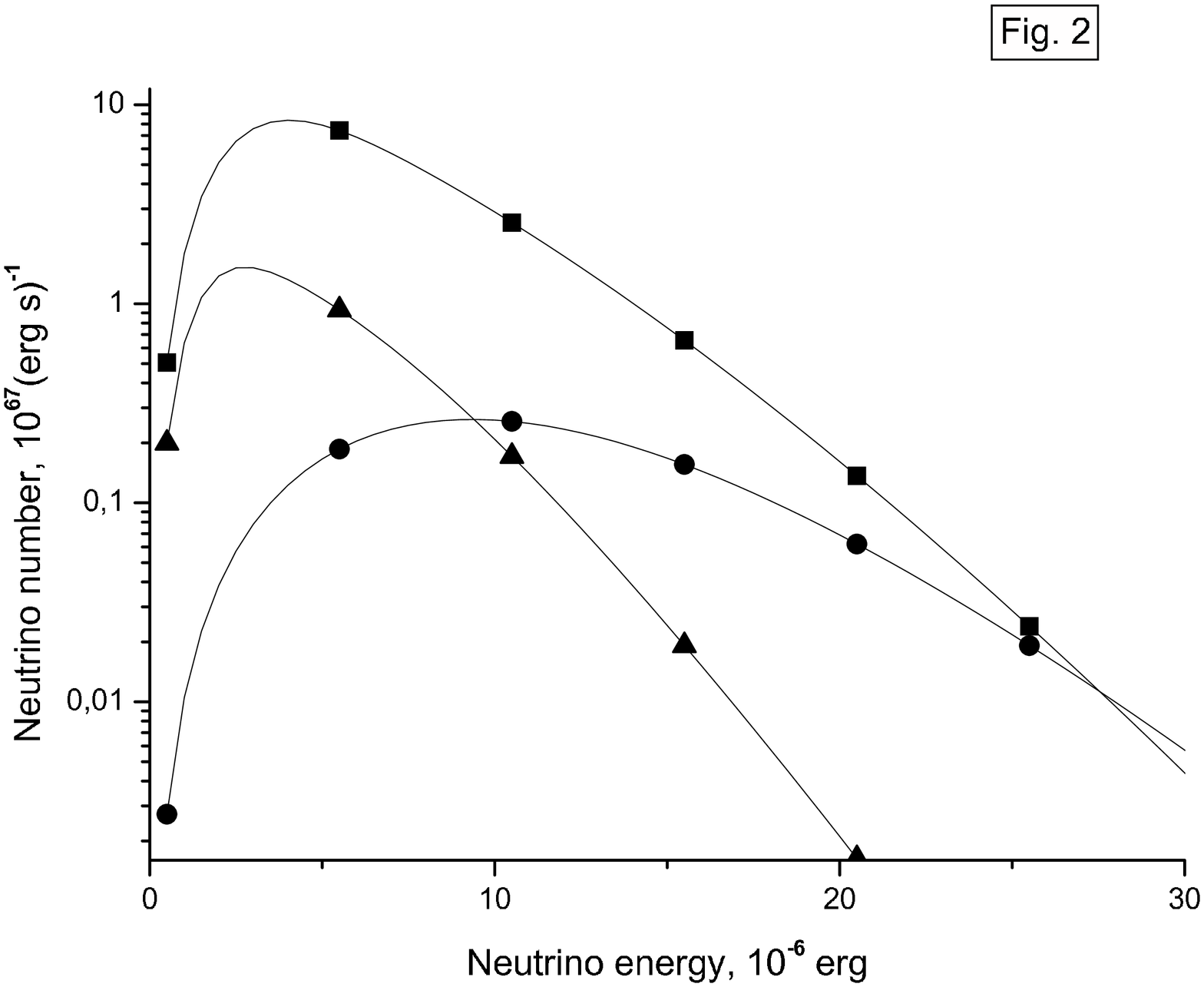}}
\caption{Spectra of the neutrinos at times $t^{*}=-10.55$~{s}
(curve with squares, luminosity $10\%$ of the maximum),
$t^{*}=-1.25$~{s} (curve with circles, maximum luminosity),
and $t^{*}=7.25$~{s} (curve with triangles, luminosity $10\%$ of
the maximum).}
\label{fig2}
\end{figure}

Thus, using expression (\ref{n12}) for the total intensity
of the radiation, we can readily show that the function
$f(q,a)$ in (\ref{a11}) is
\begin{equation}
\label{n14} f(a,q)= \frac{1.44\cdot 10^{107}}
{q \left(1+\exp\left(\dst{\frac{q}{1,6 E_0}}-2\right)\right)} a^{-9}
\end{equation}
Thus, in our case, the source function will be
\begin{equation}
\label{n16} f(q,a)=y(q) a^\beta
\end{equation}
It is straightforward to derive the form of the light
curve and resulting spectrum in this case if we make
use of the linearity of the right-hand side of (\ref{a12}).
Expression (\ref{n16}) can be rewritten as
\begin{equation}
\label{n13} f(q,a)=\int_{0}^{\infty} y(q_0) \delta(q-q_0) a^\beta
dq_0
\end{equation}
Thus, it is clear that we can obtain the spectrum
corresponding to a source function of the form (\ref{n13}) by
replacing $\lambda$ in formula (\ref{n5}) with $y(q_0)$ and integrating
over $q_0$. In our case, we obtain after straightforward
manipulation:
\begin{eqnarray}
\label{n15} \Phi(w,t^{*})= 4 \cdot 10^{71} 
w^{(\beta+\dst{\frac{7}{2}})} \int_{0}^{\infty}
\frac{{q_0}^{-(\beta+\dst{\frac{7}{2}})}}{1+\exp\left(\dst{\frac{q_0}{1,6 E_0}}-2\right)} d{\left(\dst{\frac{q_0}{E_0}}\right)}
\times \nonumber \\ \int\rho
\left(\frac{x}{m}\right)^{\beta+\frac{3}{2}} \left[\Xi(1-m
\dst{\frac{q_0}{w}})-\Xi(1-g^2 m \dst{\frac{q_0}{w}})\right]\;
d\rho \quad \left(\frac{1}{\mbox{erg}\; \mbox{s}} \right)
\end{eqnarray}
The results of computations carried out using (\ref{q2})
and (\ref{n15}) are presented in Figs.~\ref{fig1} and~\ref{fig2}. Figure~\ref{fig1}
depicts the light curve of the collapsing star, which
has a characteristic bell-shaped appearance. The duration
of the neutrino pulse is approximately $10$~s, i.e.,
$10\displaystyle{\frac{r_g}{c}}$.

Figure~\ref{fig2} presents neutrino spectra at times $t^{*}=-1.25$~{s} (curve with circles), $t^{*}=-10.55$~{s} (curve with
squares), and $t^{*}=7.25$~{s} (curve with triangles), with
the luminosity being maximum in the first case and
$10\%$ of the maximum in the latter two cases. We can
see that the spectrum becomes softer with time due to
the influence of the redshift.

\section{CONCLUSION}
The model presented here can be applied with
source functions (\ref{a11}) with fairly arbitrary shapes. Our
approximation that the shape of the spectrum of the
radiating matter does not depend on temperature in
our derivation of formula (\ref{n11}) was made only to simplify
the calculation. If the source function cannot be
presented in the form (\ref{n16}), it must be expanded in a
Loran series,
$$
f(q,a)=\sum_{n=-\infty}^{\infty}
y_n(q) a^n
$$
and the linearity of the problem taken into account;
i.e., it is necessary to obtain a solution for each term
of the series separately and then sum the results.

Thus, our solution for the properties of a neutrino
pulse arising during the collapse of a supermassive
star can be used to consider various processes that
generate neutrinos (plasma neutrinos, pair annihilation,
and so forth). There is no question that the
solution does not model a real collapse completely
accurately; in particular, the assumption that the star
is uniform during the collapse is crude. However, the
resulting solution is very simple while also displaying
all the properties that are characteristic of real systems.
This makes it possible to obtain light curves
and spectra for such stars that are close to the real
curves without having to carry out complex three-dimensional
computations.

\section{ACKNOWLEDGMENTS}
One of the authors (A. N. B.) thanks S. L. Karepov
and Yu.I. Khanukaev for useful discussions. This
work was partially supported by the Russian Foundation
for Basic Research (project codes 01-02-06146,
02-02-06596, 02-02-16900) and INTAS (INTAS-EKA
grant 99-120 and INTAS grant 00-491).

\end{document}